\documentclass[letterpaper]{JHEP3}

\usepackage{amssymb}
\usepackage{amsmath}

\preprint{{\small hep-th/0202091}} 

\title{Stringy negative-tension branes\\ and
the second law of thermodynamics}
\author{Donald Marolf\footnote{E-mail: {\tt marolf@physics.syr.edu}}  and
 Simon F. Ross\footnote{E-mail: {\tt
S.F.Ross@durham.ac.uk}}\\ ${}^{*}$ 
Physics Department, Syracuse University, Syracuse, New York 13244\\
$^\dagger$  Centre for Particle Theory, Department of
Mathematical Sciences, University of Durham, South Road, Durham DH1 3LE,
United Kingdom}
\date{February, 2002}
\abstract{Negative energy objects generally lead to instabilities and
  a number  
of other disturbing behaviors. In particular, negative energy fluxes
lead to a breakdown of the classical area theorem for black hole
horizons, which can lead to violations of the second law of
thermodynamics. The negative energy objects that arise in string
theory involve special boundary conditions which remove the
perturbative instabilities. We show that they have additional special
features which allow them to evade contradiction with the second
law. We identify one mechanism which applies for most orientifold
planes in string theory, and distinct mechanisms for the O8-plane and
the AdS soliton.
}
\begin{document}
%%%%%%%%%%%%%%%%%%%%%%%%%%%%%%%%%%%%%%%%%%%%%%%%%%%%%%
\section{Introduction}
\label{intro}

\renewcommand{\thefootnote}{\arabic{footnote}} 
\setcounter{footnote}{0}

Branes with negative tension have experienced a recent surge in
popularity.  In part, this stems from interest in string theoretic
objects called orientifolds \cite{Sag,H,Joe,GP} as perturbative string
calculations tell us that their tension is sometimes less than zero.
In addition, negative tension branes have found interesting
applications in certain brane-world scenarios (e.g., in \cite{RS} and
related work). Indeed, to the authors' knowledge all such
scenarios that address the gauge hierarchy problem contain some form
of negative tension object.

It is important to recall that negative tension branes carry negative
energy both in the sense that their total energy is less than zero and
in the sense that they violate the weak null energy
condition \cite{WaldBook}.  Mention of such objects naturally raises
the question of whether they pose a danger to one's physical theory.
On the one hand, the bold physicist may point out the orientifold
construction of string theories, relying on string theory's track
record of producing novel pieces of physics which turn out to be
surprisingly robust.  Past examples of this sort include certain kinds
of topology changing processes \cite{AGM} and dualities between
gravitating theories and non-gravitating gauge theories
\cite{Juan,ISMY}.

On the other hand, a more conservative physicist may point out
excellent reasons to be skeptical of negative energies\footnote{While
it is understood that negative energy fluxes can arise in quantum
field theory \cite{EGJ,Kuo}, there is significant evidence that such
fluxes face fundamental constraints \cite{Tipler,Romans1,FR1}
restricting them to domains where quantum mechanical effects are
important and in which their effects are more benign (see
e.g. \cite{Borde}).  In contrast, as will be discussed in detail in
section \ref{others}, the negative energies mentioned above can be
relevant in purely classical regimes.}.  After all, unbridled negative
energies generate disturbing dynamical instabilities.  In addition,
when gravitational effects are considered, unconstrained negative
energies allow the formation of traversable wormholes
\cite{MorTh,MTY}, ``faster than light travel'' \cite{Alc,Olum,GW},
naked singularities \cite{FR2,FR3}, and closed timelike curves (aka
time machines) \cite{MTY,Ev}.  Most importantly for the present work,
the foundation of black hole thermodynamics rests on the classical
area theorem \cite{HE} which states that, {\em if the energy of matter
is non-negative}, the area of black hole horizons cannot decrease. In
the presence of negative energies, the entropy associated with black
holes can decrease, and the conservative physicist would suspect that
this will lead to violations of the second law of thermodynamics.

Clearly then, it is important to reconcile the bold and conservative
points of view and to discover what (if any) special properties
particular negative energy objects may possess that render them immune
to such concerns.  To date, discussions in the literature have
concentrated on the fact that orientifolds have `orbifold boundary
conditions,' meaning that the perturbative description of an
orientifold involves a quotient of some smooth spacetime by a discrete
$Z_2$ symmetry.  For an orientifold $p$-plane, this is a $Z_2$
symmetry that leaves a $(p,1)$ spacetime surface invariant and
reflects the other directions about this surface.  Note that this
quotient implies that the spacetime fails to be asymptotically flat in
the usual sense and that the change of boundary conditions forbids
pair production of orientifolds from the vacuum\footnote{Here we refer
to orientifold points and non-compact or non-contractible
orientifolds. Presumably, contractible orientifold planes must have
positive energy to avoid this problem since their presence is
compatible with the usual boundary conditions.  This is a plausible
result, though we are not aware of any literature on the subject.}.
Certain other important aspects of orientifold physics can be captured
by working on the two-fold covering space and requiring the physics to
be $Z_2$ symmetric.  In particular, this removes the perturbative
negative energy oscillations of the $p$-plane.  Since the $Z_2$
quotient resolves at one stroke two potential instabilities of concern
to our conservative physicist, it provides much encouragement to bold
physicists wishing to use such objects in brane-world scenarios.

While it may be tempting to regard the creation of closed time-like
curves, wormholes, and naked singularities as intriguing possibilities
rather than physical inconsistencies, few physicists would be willing
to compromise the second law of thermodynamics.  Indeed, a famous
quote from Albert Einstein \cite{AE} states his belief that
thermodynamics was the most likely element of physics to survive
without change -- in particular, that our understanding of
thermodynamics is more reliable than either that of quantum mechanics
or relativity.  Presumably, it is also more reliable than our
understanding of string theory.

It was argued in \cite{MT} that the orbifold boundary condition alone
is not sufficient to enforce the second law for negative energy
branes.  We should first remark that although one can in principle
violate the second law using negative energy and ordinary matter by,
for example, throwing negative energy into a star and thereby lowering
its entropy and temperature, such concerns are in fact removed {\it
for ordinary matter} if the negative energy is confined to an orbifold
plane.  The association of the negative energy with the unusual
boundary conditions makes it impossible for the orbifold to be
destroyed by ordinary matter and thus forbids its negative energy from
thermalizing and lowering the temperature of our star.

On the other hand, the issue becomes much more subtle when black holes
are involved.  Unlike normal matter, black holes absorb the energy of
any object that passes inside them. They also have the capability to
hide the would-be orbifold plane.  The usual tool for considerations
of black hole thermodynamics is the Raychaudhuri equation
\cite{WaldBook}, which shows that a negative energy object crossing
the horizon of a black hole in a region of negligible shear
necessarily causes null rays to defocus, so that the horizon area (and
thus, at first sight, the entropy) tends to decrease.  While one does
not typically describe objects with orbifold boundary conditions as
`moving,' it is straightforward to instead consider a black hole that
is thrown toward our negative tension brane.  The resulting collision
is the same in either case, and corresponds to a symmetric collision
of two black holes with a negative tension object in the covering
space.  The second black hole complicates the analysis somewhat but,
depending on the co-dimension $q$ of the negative tension brane, a
scaling analysis shows that choosing the black holes to be either very
large (for $q=1$) or very small (for $q > 2$) compared to the
gravitational length scale of the negative tension brane makes the
mass of this second black hole negligible in comparison with the
amount of negative energy absorbed from the brane.  In such regimes,
the second black hole should not interfere strongly with the argument
above.

In addition to the concern that the negative energy object will cause
defocusing of light rays and produce a local decrease in the horizon
area, one might have the simpler concern that absorbing a negative
energy object would reduce the mass of the black hole so that the
final equilibrium configuration would be expected to have smaller
horizon area than the initial one. This is essentially a repeat
performance of our star argument above, with the notable different
that the black hole should in fact be able to thermalize the negative
energy of an orientifold.  However, one can easily see that this
concern is groundless in most cases. A negative energy object is
gravitationally repulsive, so that in order to collide, the brane and
black hole must be impelled toward each other with a finite kinetic
energy\footnote{That this kinetic energy is necessarily positive is
most easily seen by considering the negative tension brane to be
unmoving and thinking of the kinetic energy as associated with the
relative motion of our black hole and its image in the two-fold
covering space.}.  Since the energy of any object vanishes when held
static at a black hole horizon, this kinetic energy must be as large
as the rest mass of the negative tension brane in order for the brane
to just reach the horizon of the black hole.  In general then, the
collision occurs only if the kinetic energy is large enough to
counterbalance the negative contribution to the mass of the final
object.  While there remains a concern about black hole entropy during
the collision itself, one therefore expects the final equilibrium
state to have larger energy and thus larger entropy than did the
initial state.

When the negative tension brane has co-dimension 1 (as considered in
\cite{RS} and \cite{MT}), however, this argument breaks down.  Such
branes are gravitationally attractive.  In these cases, one needs to
address the question of the entropy of the final state. The O8 plane
and AdS soliton provide examples of this type; they are discussed in
section \ref{o8} and \ref{soliton}.

The authors of \cite{MT} explored collisions between a black hole and
a negative tension brane in a low dimensional toy model.
Interestingly, they found that entropy considerations became moot as
such collisions necessarily gave rise to a spacelike (big-crunch like)
singularity that engulfed the entire space and prevented any notion of
a final equilibrium configuration.  Nevertheless, one intuitively
suspects that this non-perturbative dynamical instability is linked to
the above concerns about the second law.  It is interesting to note
that the instability of \cite{MT} remains when the system is embedded
(as is straightforward to do\footnote{We thank Jon Bagger and Raman
Sundrum for this observation.})  in a supersymmetric system of the
sort considered in \cite{SUSY}.  An investigation is currently in
progress \cite{MRST} of whether the instability might be an artifact
of the low dimensional setting used in \cite{MT}, but preliminary
results for a 2+1 dimensional negative tension brane in $AdS_{3+1}$
indicate that the instability will survive in higher
dimensions \cite{Pri}. 

Such a dynamical instability would be as undesirable
for physical models as violations of
the second law. We are therefore compelled to seek an alternative
resolution of the problems identified above for the stringy negative
tension objects.  We will see that this study uncovers a new and
perhaps surprising property of (most) orientifolds that removes any
concern about violations of black hole thermodynamics.  This property is related to an
unusual set of couplings between the gravitational field and massive
modes associated with various classes of non-perturbative stringy
effects.  Because they involve the curvature, such couplings guarantee
that black hole entropy does not reduce strictly to the horizon
area \cite{wald} for a black hole in the neighborhood of an
orientifold.  This allows the entropy of a black hole to increase even though its area
decreases as the horizon encounters a
negative tension brane.

For each known case of a perturbatively stable stringy negative energy
object, we find some novel mechanism that protects the system from
violations of the second law.  In most cases, this mechanism is the
one just mentioned.  However, the D8-brane and the AdS soliton \cite{hor-myers} (which
has already been used in brane-world constructions \cite{LMW})
constitute interesting exceptions.  For each of them, we uncover a
distinct mechanism for consistency with the second law.  One suspects
that stable consistent negative energy objects can exist only when
they have similarly clever properties and a natural first conjecture
is that these three examples provide a complete list.  One therefore
expects that while more simplified phenomenological descriptions of
negative tension objects (e.g., of the sort used in \cite{RS}) can
point the way to intriguing and useful new phenomena, they remain at
best a rough approximation to the physics involved and when studied in
detail are likely to have instabilities of the sort described in
\cite{MT,MRST}.

We begin our analysis in section \ref{o6} with the case of the
orientifold six-plane (O6-plane).  This case has the advantage of a
known strong coupling description as a purely gravitational soliton
\cite{Sen,Seiberg,SW} (the Atiyah-Hitchin manifold \cite{AHbook}) in M
theory.  In such a manifestly classical setting it is straightforward
to explore issues of black hole entropy and non-perturbative
gravitational couplings in detail.  As we will describe, at strong
coupling the ten-dimensional description of the O6-plane necessarily
involves not only the massless fields of type IIA supergravity but
also the 11-dimensional Kaluza-Klein modes, which in ten-dimensions
are described by an infinite tower of fields at integer multiples of
the D0-brane mass.  These fields are associated with the well known
`exponential corrections' that deform the negative mass Euclidean
Taub-NUT manifold \cite{HP} to the smooth Atiyah-Hitchin manifold, and
which can be seen in a field theoretic treatment of a D2-brane probe
near an O6-plane. We argue that there is clearly no violation of the
second law in the M theory description, so curvature couplings to
these non-perturbative corrections resolve the puzzle as outlined
above.

We then consider other negative energy orientifolds in section
\ref{others}.  This section begins with a scaling analysis to
determine whether one would expect these objects to be
subject to similar effects.  The answer is in the affirmative for most
cases, and we find in particular that for the strongly coupled
O5-plane they must again come from fields at a non-perturbative mass
scale (in this case, the scale set by the D-string tension).  We argue
that such effects are associated with the onset of a
Kosterlitz-Thouless transition \cite{KT} in the field theory of a
D-string probe near an O5-plane.  At strong coupling, the associated
instanton/anti-instanton pairs are unbound and produce
non-perturbative corrections to the corresponding moduli space metric.

We turn in section \ref{o8} to the O8-plane, which forms the single
exceptional orientifold case.  Instead of relying on non-minimally
coupled massive fields, the O8-brane turns out to avoid difficulties
through the requirement that it appear in the vicinity of positive
tension D8-branes.

A final negative energy object is the ADS soliton, which we address in
section \ref{soliton}.  This can be considered either as a purely
gravitational negative energy soliton or, to a certain extent, as a
negative energy brane in one lower dimension.  Despite the similarity
to the O6-plane and its Atiyah-Hitchin description in M-theory, this
negative tension brane avoids thermodynamic difficulties through a completely
different effect.  We will see that it is simply impossible for such a
negative tension brane to cross the event horizon of a black hole!
Interestingly, this effect is necessarily associated with a vanishing
(Einstein frame) worldvolume metric on the lower-dimensional brane,
again leading to a breakdown of a description in terms of pure Einstein-Hilbert gravity
coupled to a negative tension brane.
See \cite{LMW}, however, for how to use such solitons to address the hierarchy problem in 
brane-world settings.

\section{The orientifold six-plane}  
\label{o6}   

For most orientifolds, we wish to argue that the spacetime
description receives corrections which break the relationship between
entropy and area for nearby black holes. In this section, we will explore in detail
the case of an O6-plane\footnote{In particular, the O$6^-$-plane in the notation of
\cite{BGS}.} in the strongly-coupled IIA theory. The effective
description at low energies is then eleven-dimensional supergravity
compactified on a circle of radius $R_{11} \sim g l_s$ where $g$ is the string coupling and
$\ell_s$ is the string length.  In this case, one 
expects the dominant corrections to the ten-dimensional geometry to
come from the massive Kaluza-Klein modes associated with this
compactification (and which become light in the strong coupling limit). 
The advantage of focusing on this case is that non-perturbative effects associated with $R_{11}$
are fully encapsulated in the eleven-dimensional supergravity action.  Thus, we
gain a much more detailed understanding of the effects of the
corrections than in cases where the dominant corrections come from,
say, massive string modes. 
Note that because adding
additional orientifolds changes the boundary conditions, one cannot
suppress corrections by considering an arbitrary number of coincident
O6's, as one can with D-branes.

Recall that the tension of an O6-plane is 
\begin{equation}
\tau_6 = - {2 \over (2\pi)^6 g {\ell_s}^{7}},
\end{equation}
so that the radius of curvature of the ten-dimensional metric 
is of order
\begin{equation}
R \sim G_{10} \tau_6 \sim g l_s, 
\end{equation}
where $G_{10}$ is the ten-dimensional Newton's constant. Hence, $R$ is
of order $ R_{11}$, but $R$ is much larger than $l_s$, as well as the
length scales $l_p^{(10)} = g^{1/4} l_s$ and $l_p^{(11)} = g^{1/3}
l_s$ associated with the ten- and eleven-dimensional Planck lengths.
Thus, while Kaluza-Klein corrections are relevant and an
eleven-dimensional description is needed to capture the full physics,
no further corrections need be considered.

Below, we first review how the eleven-dimensional geometry can be
obtained by stringy methods (section \ref{probe}) and then describe
the geometry itself (section \ref{AH}) which provides the resolution
of our concerns from the eleven-dimensional perspective.  We then
discuss the entropy (section \ref{10d}) and energy
(section \ref{mass-stress}) from the ten-dimensional point of view.

\subsection{D2-brane probe calculation}
\label{probe}

We now briefly review how
to obtain the complete eleven-dimensional geometry
from a probe D-brane
calculation~\cite{Seiberg,SW,Dorey,HP}, though we refer the reader 
to the original references for further details. 

Consider a probe D2-brane with its worldvolume aligned parallel
to two of the directions in the orientifold plane. A D2-brane in the
orientifold background will carry a worldvolume field theory with
$SU(2)$ gauge invariance and $N=4$ supersymmetry in $2+1$
dimensions. There are seven scalar fields which parametrize the
directions orthogonal to the D2-brane in the ten-dimensional geometry
which split naturally into four fields corresponding to the directions
along the O6-plane (which decouple from the other fields) and three
fields $\phi_i, i=1,2,3$ (corresponding to the overall transverse
space). These scalars transform in the adjoint of $SU(2)$, and their
potential energy is minimized when they all commute. At a generic
minimum of the potential, the gauge group is broken to $U(1)$ and,
since we are in 2+1 dimensions, this massless gauge field can be
dualized to give an additional scalar $\sigma$. Since it arises from a
$U(1)$ gauge field, $\sigma$ is periodically identified, and $\sigma$
in fact corresponds to the M-theory circle. We therefore have a
four-dimensional Coulomb branch for this field theory for which the
metric on moduli space (cross the seven flat directions along the
O6-plane) is just the ten-dimensional spatial geometry of the
O6-plane\footnote{Here we assume that the three-form potential
vanishes and that the eleven-dimensional geometry is of the form ${\mathbb
R} \times {\cal M}_{10}$ (a direct product of metrics) where ${\mathbb R}$
is the time direction.  This assumption will be justified a posteriori
by the fact that the result satisfies the eleven-dimensional
supergravity equations of motion.}.

At the classical field theory level, the metric on the space of
$\phi_i, \sigma$ is just the flat metric on $(\mathbb{R}^3 \times
S^1)/\mathbb{Z}_2$, where the $\mathbb{Z}_2$ identification arises
from the action of the Weyl group of $SU(2)$.  Supersymmetry
guarantees that the moduli space cannot be lifted by quantum
corrections and that the metric on moduli space remains
hyper-K\"ahler. By studying the action of the symmetries one can show
that the one-loop contribution generates a cross term between the
$\phi_i$ and $\sigma$. In terms of the ten-dimensional string theory,
this correction corresponds to the Ramond-Ramond charge carried by the
O6-plane. The hyper-K\"ahler structure implies that there are no new
corrections at higher orders in perturbation theory, but there may be
non-perturbative corrections.  Indeed, one can identify instantons
associated with D0-brane exchange between the probe and the
orientifold that yield an infinite series of corrections exponentially
suppressed by powers of the D0-brane mass.  Since there is no Higgs
branch for this theory, the metric on the Coulomb branch must be
non-singular~\cite{SW}; together with the supersymmetry and asymptotic
structure, this is enough to determine the metric uniquely. The
appropriate metric is the Atiyah-Hitchin metric of \cite{AHbook}.
Thus, the flat metric receives a one-loop correction which accounts
for the effects of the bulk Ramond-Ramond gauge field, and instanton
corrections which remove the singularity at the origin to give the
Atiyah-Hitchin metric. The eleven-dimensional metric for the O6-plane
is then this metric cross a flat $\mathbb{R}^{6,1}$.

\subsection{The Atiyah-Hitchin metric}
\label{AH}

The Atiyah-Hitchin metric is a non-singular $SO(3)$ symmetric
hyper-K\"ahler manifold, which also appears in the study of a
two-monopole moduli space. The metric takes the form~\cite{monopole}
\begin{equation} \label{ah}
ds^2 = {b^2 \over r^2} dr^2 + a^2 \sigma_1^2 + b^2 \sigma_2^2 + c^2
\sigma_3^2, 
\end{equation}
where
\begin{eqnarray}
\sigma_1 &=& - \sin \psi d\theta + \cos \psi \sin \theta d\phi,
\nonumber \\ 
\sigma_2 &=& \cos \psi d\theta + \sin \psi \sin \theta d\phi,
\nonumber \\
\sigma_3 &=& d \psi + \cos \theta d\phi,
\end{eqnarray}
and $a,b,c$ are functions only of $r$. The ranges of the coordinates
are $0 \leq \theta \leq \pi$, $0 \leq \phi \leq 2\pi$, $0 \leq \psi
\leq 2\pi$, and $\pi R_{11}/2 \leq r < \infty$. In addition, an
identification is imposed under the $\mathbb{Z}_2$ symmetry
\begin{equation}
\theta \to \pi - \theta, \quad \phi \to \phi + \pi, \quad \psi \to
-\psi.
\end{equation}
At large $r$, the functions $a,b,c$ take the form
\begin{eqnarray}
a &=& r \left( 1 - {R_{11} \over r} \right)^{1/2} 
- \frac{8r^2}{R_{11}}\left(1 - \frac{R_{11}^2}{8r^2}
\right) e^{-\frac{2r}{R_{11}}} + {\cal O}(e^{-\frac{4r}{R_{11}}} ), 
\nonumber \\
b &=& r \left( 1 - {R_{11} \over r} \right)^{1/2} 
+ \frac{8r^2}{R_{11}}\left(1 + \frac{R_{11}}{r} - \frac{R_{11}^2}{8r^2}
\right) e^{-\frac{2r}{R_{11}}} + {\cal O}(e^{-\frac{4r}{R_{11}}} ), 
\nonumber \\ 
c &=& - R_{11} \left( 1 - {{R_{11}} \over r} \right)^{1/2} 
+ {\cal O}(e^{-\frac{2r}{R_{11}}} ), 
\end{eqnarray}
where $2\pi R_{11}$ is the asymptotic circumference of the $S^1$.
The above two features together imply that the geometry asymptotically
approaches the
flat metric on $(\mathbb{R}^3 \times S^1)/\mathbb{Z}_2$, with $\psi$ the
coordinate on the $S^1$.  Thus, it is the angle $\psi$ that
corresponds to the M-theory circle and the
reduction on this circle 
to ten dimensions gives a
magnetic charge under the Kaluza-Klein gauge field. 

The exponentially
small corrections at large $r$ represent the effect of the instantons
described in the preceding section. Because of these
corrections, the metric closes off in a regular fashion at $r=\pi R_{11}/2$,
where $b = -c = \pi$, and $a \to 2r-\pi R_{11}$. There is
no region with $r < \pi R_{11}/2$.  For more details,
see~\cite{monopole}.

For our purposes, the important point is that the eleven-dimensional
geometry is a completely smooth vacuum solution; the above metric is
Ricci flat.  Despite the vacuum geometry, this spacetime actually has
a negative ADM mass (tension).  The failure of the positive mass
theorems \cite{SY,WPM} is associated with the unusual boundary
conditions which, for example, rule out the existence of the
asymptotically constant spinors used in the Witten proof \cite{WPM}.
The complete description of the collision of a black hole with the
O6-plane is then given by a black string wrapped on the compact
direction in 11d `moving in the Atiyah-Hitchin space'{}\footnote{Such
language is strictly appropriate only for a perturbatively small black
string, but it is convenient to use this phrase to refer to large
black strings as well.}. Since this spacetime satisfies the weak
energy condition --- indeed, there is no local stress-energy of any
kind --- in the 11d description, the usual 11-dimensional area theorem
tells us that the area, and hence entropy, of this black string
solution is non-decreasing.  Thus, once we take the exponential
corrections fully into account, the apparent violation of the second
law is completely resolved.

\subsection{Entropy from the ten-dimensional point of view}
\label{10d}

The eleven-dimensional picture removes concern about the second law
and provides a very satisfying result.  However, it is useful to
explore the mechanism by which the second law is saved from the
ten-dimensional point of view. To flesh out the details, we now
re-examine the Atiyah-Hitchin metric from the ten-dimensional point of
view and see how the na\"\i ve argument in the introduction is
invalidated.

First of all, we note that the Atiyah-Hitchin metric (\ref{ah})
depends on $\psi$, so that $\partial_\psi$ is not a Killing vector
field. This implies that the dimensional reduction to ten dimensions
will contain excited massive fields from the Kaluza-Klein modes.  We
wish to argue that the presence of these massive modes implies that
the usual relation between the entropy and the area of the black hole
horizon is modified. The essential point is that the formula $S= A/4$
is valid only in Einstein gravity minimally coupled to matter
fields. Wald and Iyer undertook a detailed study of a family of
actions with further curvature terms, and showed that the first law of
black hole thermodynamics requires the definition of the entropy to be
suitably generalized~\cite{wald}.  The full entropy expression is
obtained by differentiating the Lagrangian with respect to its
dependence on the Riemann tensor, so that each curvature coupling
provides a separate contribution to the entropy of any black hole.

In our case, the presence of the massive modes will produce such additional
curvature terms. In general, the ten-dimensional action
involving these massive modes is a complicated non-linear
expression. However, these couplings become tractable when the massive modes
can be treated as a small perturbation.  This will be enough for our purposes
as we seek only a qualitative understanding
in ten-dimensional terms.

Writing the 11d metric as $g_{ab} = g^0_{ab} +
h_{ab}$, the quadratic action for the perturbation $h_{ab}$
is~\cite{action}
\begin{equation}
\label{quad}
S = {1 \over 32 \pi G} \int d^{11} x \sqrt{-g^0} \  h^{ab} A_{abcd} h^{cd},
\end{equation}
where $g^0_{ab}$ has been used to raise the indices and
\begin{eqnarray}
\label{2nd}
A_{abcd} &=& {1 \over 4} g^0_{cd} \nabla_a \nabla_b - {1 \over 4} g^0_{ac}
\nabla_d \nabla_b + {1 \over 8} (g^0_{ac} g^0_{bd} + g^0_{ab} g^0_{cd})
\nabla_e \nabla^e + {1 \over 2} R^0_{ad} g^0_{bc} - {1 \over 4} R^0_{ab}
g_{cd} \nonumber \\ &&+ {1 \over 16} R^0 g^0_{ab} g^0_{cd} - {1 \over 8} R^0
g^0_{ac} g^0_{bd} + (a \leftrightarrow b) + (c \leftrightarrow d) + (a
\leftrightarrow b, c \leftrightarrow d).
\end{eqnarray}
Here $R^0_{ab}$ ($R^0$) denotes the Ricci tensor (scalar) of the
background spacetime $g_{ab}^0$. Similarly, $\nabla$ is the covariant
derivative associated with $g^0_{ab}$.  We are mostly interested in
the curvature couplings, which can be rewritten as
\begin{equation}
\label{ct}
S_{curv} = {1 \over 32 \pi G} \int d^{11} x \sqrt{-g^0} R^0_{ab}
\left(2 h^{ac} h_c{}^b - 
h^{ab} h_c{}^c + {1 \over 4} h_c{}^c h_d{}^d - {1 \over 2} h^{ab}
h_{ab} \right).
\end{equation}

We  take the background $g^0_{ab}$ to be independent of
$\psi$ and expand
$h_{ab}$ in Fourier modes, 
\begin{equation}
h_{ab} = \sum_{m \neq 0} h_{ab}^{(m)} e^{im \psi}.
\end{equation}
Note that we have taken the perturbation of the zero-mode to vanish,
as the perturbation is to represent only the massive fields.  It is
for this reason that terms in the action that are linear in $h_{ab}$
integrate to zero regardless of whether $g^0_{ab}$ is taken to be a
solution of the equations of motion.  Reduction of (\ref{ct}) to
ten-dimensions is straightforward once a few conventions are fixed.
We choose to work with what we call the ``na\"\i ve Einstein metric''
$g^{10,E}$ (and the associated dilaton $\phi$ and 1-form $A_a$) in
ten-dimensions which is given by the usual algebraic expressions in
terms of $g^0_{ab}$:

\begin{equation}
\label{kkred}
g^0_{ab} = e^{-\phi/6} g^{E,10}_{ij} dx^i dx^j + e^{4\phi/3} (d\psi + A_i dx^i)^2.
\end{equation}
Here we use $i,j$ to represent ten-dimensional indices.  We use the
phrase `na\"\i ve Einstein metric' because the associated action is
readily seen to contain the curvature couplings

\begin{eqnarray}
S_{curv} &\sim& \int d^{10} x e^{-\phi/6} 
\sqrt{-g^{E,10}} R_{ij}^{(E,10)} x^i_{,a} x^j_{,b}  \nonumber \\
&\times& \sum_m \left(2 h^{ac(m)}
h_c{}^{b(m)}  -
h^{ab(m)} h_c{}^{c(m)} + {1 \over 4} h_c{}^{c(m)} h_d{}^{d(m)} - {1
\over 2} h^{ab(m)} h_{ab}^{(m)} \right),  
\end{eqnarray}
where $R_{ij}^{(E,10)}$ is the Ricci tensor of $g_{ij}^{E,10}$ and the matrix of derivatives
$x^i_{,a}$ implements the pull back of ten-dimensional forms to eleven dimensions.

While this calculation is only valid when the massive modes are a
small perturbation, we wish to suggest that similar considerations
apply when the massive modes are of order one, as in the region near
$r=\pi R_{11}/2$ in the Atiyah-Hitchin metric.  In particular, we
suggest that the terms in (\ref{quad}) and the corresponding higher
corrections will continue to source important curvature terms in the
ten-dimensional action, which will produce order-one corrections to
the entropy formula.  The full entropy must after all give just the
eleven-dimensional area of the black hole and there is no reason in
the presence of massive excitations for this to agree with the area as
measured in ten-dimensions.

Let us therefore summarize our picture of the collision of a black
$p$-brane and an orientifold plane.  When the black $p$-brane is
absorbing the orientifold plane, massive Kaluza-Klein modes will be
excited near the horizon. What happens to these modes subsequently? If
the black $p$-brane completely engulfs the orientifold plane, we would
not expect the black $p$-brane to be capable of supporting `massive
hair'. In the linearized analysis, it is easy to see that the massive
modes must die off exponentially, so long as the black $p$-brane has a
radius $r_0$ satisfying $r_0 \gg R_{11}$ (so that it is a truly
ten-dimensional solution). This is because in eleven dimensions we are
considering just perturbations of the black string of the form $h_{ab}
\sim e^{im\psi}$. But these are the perturbations considered by
Gregory and Laflamme~\cite{GL}, who showed that they will decay
exponentially in time for all $m > m_* \sim R_{11}/r_0$. Thus, the
massive modes are absorbed by the black $p$-brane, which settles down
to a regular solution carrying just the magnetic Kaluza-Klein gauge
charge.

{}From the ten-dimensional point of view, the picture we want to suggest
is thus that the orientifold will first fall across the 
event horizon, lowering the area of the black $p$-brane. However, the
massive Kaluza-Klein fields will then have order-one excitations in the vicinity of the horizon,
so the area will differ substantially from the entropy. These massive modes will
then be absorbed by the black $p$-brane.  Presumably, this influx carries positive energy into
the black brane and raises its area so that the
final state will be a black brane with energy and horizon area greater
than the initial one. Since the massive modes die away, the ten- and eleven-dimensional
areas must once again agree in the final state.  One can then see that the final area is greater
from the eleven-dimensional picture where there is no negative
energy.  Alternatively, one can use the general argument in the introduction that the
kinetic energy must be at least big enough to overcome the negative
energy from the orientifold tension.

\subsection{Energy from the ten-dimensional point of view}
\label{mass-stress}

We have argued that the description of the horizon dynamics should be
rather different from the 10- and 11-dimensional points of view, with
the 10-dimensional horizon shrinking during the collision while the 11-dimensional horizon
grows.  We then suggested that this difference can be traced to the
massive modes.  In section \ref{10d} above we showed that these massive
modes do induce a series of perturbative corrections to black hole entropy.
We now verify the flip-side of our suggestion by showing that the massive
modes lead to perturbative negative energy fluxes from the ten-dimensional point of view
and so can be associated directly with the defocusing of null rays and the shrinking
of black hole horizons.  In particular, we argue that at finite
coupling the negative energy is not confined to an infinitely thin brane, but instead spreads
out over a finite region of space.   We again study
these excitations in the perturbative regime where
there is a clear distinction between the massive fields and the
massless IIA fields.  For the solution (\ref{ah}), this corresponds to the region 
$r \gg R_{11}$.

We wish to show that Kaluza-Klein reduction of (\ref{ah}) to
10-dimensions yields massive modes which
violate the weak null energy condition ($k^ak^bT_{ab} >
0$) for some null vectors $k$.  Here, we speak
of the stress-energy tensor associated with the na\"\i ve Einstein metric $g^{E,10}$ introduced
above.  

If the massive fields were the only sources in ten-dimensions, the sign of the massive mode
energy would be related by the Einstein equations to 
the sign of $R^{E,10}_{ab} k^ak^b$,
which must be positive to prove the usual black hole area
theorem.  In the present setting, the massless ten-dimensional matter fields (the dilaton and
the Ramond-Ramond 1-form) are also excited and, in fact, dominate over the massive modes at
large $r$.  The stress-energy tensor of these massless fields satisfies the weak
energy condition, and therefore so must the full stress-energy tensor in any
perturbative treatment.  However, our goal here is to analyze the contribution of the
massive modes $T^{massive}_{ij}$
in particular, which can be obtained
by simply subtracting the massless contribution $T^{massless}_{ij}$ from
the full stress-energy tensor $T^{full}_{ij}$.  That this difference has
a simple representation in terms of eleven-dimensional quantities is readily seen
by using the ten-dimensional Einstein equations (which include the effects of the massive
modes) to express 
$T^{massive}_{ij}$ in terms of a variation of the massless IIA action:
\begin{eqnarray}
k^ik^j T^{massive}_{ij} &=& k^ik^j T^{full}_{ij} - k^ik^j T^{massless}_{ij} = 
\frac{1}{8 \pi G_{10}}  k^ik^j R^{(E,10)}_{ij} - k^ik^j T^{massless}_{ij} \nonumber \cr 
&=&  \frac{2}{
\sqrt{- g^{E,10}}}
k^i k^j
\frac{\delta}{\delta g^{(10,E)ij}} S_{massless \ IIA}
= \int d\psi \frac{e^{-\phi/6} 
\sqrt{-g^{11} } } { 16 \pi^2 G_{10} \sqrt{- g^{10,E}}}
k^i k^j R^0_{ij},
\end{eqnarray}
where we have assumed that $k$ is null with respect to both the ten- and eleven-dimensional
metrics ($g^{E,10}_{ab}$ and $g^0_{ab}$) and the symbol $R^0_{ab}$ refers
to the Ricci tensor of the eleven-dimensional zero-mode
metric $g^{0}_{ab}$.  

As a result, we need only analyze the sign of 
$ \int d\psi k^a k^b R^{0}_{ab}$ for an appropriate null vector $k$.
Note that although the full eleven-dimensional metric $g_{ab} = g^0_{ab} + h_{ab}$
is Ricci flat in the Atiyah-Hitchin case, the zero mode piece $g^0_{ab}$ need not be and 
$k^a k^b R^{0}_{ab}$ need not vanish.
The setting here is
similar to that often used to treat gravitational radiation in which we
separate out a gravitational perturbation $h_{ab}$ and choose not to regard it not as
part of the metric (represented by $g^0_{ab}$) but instead as a `matter field' which in turn acts
as a gravitational source.

Nonetheless, the vanishing of the full Ricci tensor $R^{(11)}_{ab}$ 
is still quite useful.  In particular, 
we can expand $R^{(11)}_{ab}$ as
\begin{equation}
0 = R^{(11)}_{ab}  =
R^{0}_{ab}   + 
R^{0,1}_{ab} +
R^{0,2}_{ab} +  ...,
\end{equation}
where $R^{0,1}_{ab}$ and $R^{0,2}_{ab}$ are respectively linear and quadratic in the perturbation
$h^{ab}$.
Since $h^{bc}$ has no zero-mode part, the integral of $R^{0,1}_{ab}$ over $\psi$ vanishes and to second
order in the perturbation we have
\begin{equation}
\int d\psi  k^ak^b R^{0}_{ab} = - \int d\psi  k^ak^b R^{0,2}_{ab}. 
\end{equation} 

We may then consult \cite{WaldBook} to find\footnote{The reference \cite{WaldBook}
contains the result for perturbations around Minkowski space.  However,  variations
of the Ricci tensor can be written in terms of only the metric perturbation $h_{ab}$,
its covariant derivatives, and the unperturbed Ricci tensor.    Thus, since terms
involving $R^{0}_{ab}$ and two $h$'s are higher order, 
$R^{0,2}_{ab}$ can be obtained from the flat space result by simply replacing partial derivatives with
covariant derivatives and checking the ordering of the second covariant derivatives
against (\ref{2nd}) above.} the relation
\begin{eqnarray}
R^{0,2}_{ab} = \frac{1}{2} h^{cd} \nabla_{(a} \nabla_{b)} h_{cd}
- \frac{1}{2} h^{cd} \nabla_{a} \nabla_c h_{bd} 
- \frac{1}{2} h^{cd} \nabla_{b} \nabla_c h_{ad} +
\frac{1}{4} (\nabla_a h_{cd}) \nabla_b h^{cd} \cr
+ (\nabla^d h^c_b) \nabla_{[d} h_{c]a}
- \frac{1}{2} \nabla_d (h^{dc}\nabla_c h_{ab})
- \frac{1}{4}(\nabla^c h) \nabla_c h_{ab}
- (\nabla_d h^{cd} - \frac{1}{2} \nabla^c h) \nabla_{(a}
h_{bc)},
\end{eqnarray}
where we have used the fact that terms involving $R^0_{ab}$
are higher order in $h_{ab}.$
Here indices 
are raised and lowered with the zero-mode metric $g^0_{ab}$ and 
$h = g^{0,ab} h_{ab}$ is the trace of the perturbation in this metric.

It is now straightforward to evaluate $k^ak^bR^{0,2}_{ab}$ for various
null vectors $k$ using the large $r$ expansion of (\ref{ah}).  We
choose $k = k^t \frac{\partial}{\partial t} \pm k^r
\frac{\partial}{\partial r}$, which is clearly relevant to the horizon
of an approaching black hole.  The reader may check that such $k$ are
null simultaneously for both $g^{E,10}_{ab}$ and $g^0_{ab}$.  If we
introduce
\begin{equation}
\Delta^2 \equiv b^2 - a^2 = \frac{2^7r^3}{R^3_{11}} e^{-2r/R_{11}} \left(1
+ {\cal O}( \frac{r}{R_{11}}) \right),
\end{equation}
a somewhat tedious calculation yields
\begin{equation}
\frac{1}{2\pi} \int d \psi k^a k^b R^0_{ab} =
- (k^r)^2
\frac{3 R_{11}^2}{32r^4} \Delta^4 \left(1  + {\cal O}(\frac{r}{R_{11}}) \right) < 0. 
\end{equation}
As a result, the stress-energy tensor of the massive modes does indeed
violate the weak energy condition in our na\"\i ve ten-dimensional
Einstein frame.

Of course, this does not mean that an approaching black hole begins to
contract while the orientifold is still far away.  The energy in the
large $r$ regime is is dominated by the massless modes and must
therefore be positive.  However, we expect the massive modes to
dominate once $r \sim R_{11}$, and there is no reason to expect the
sign of the energy carried by such modes to change at that point.
Thus, the calculation above indicates that at finite coupling the full
stress-energy violates the weak energy condition over a region of size
$R_{11}$.  When the black hole encounters this region, its horizon
will begin to contract.

\section{Other Orientifold-planes}
\label{others}

We now want to consider the other Orientifold-planes of type II string
theory.  These may be classified in much the same manner as the
familiar positive tension branes in terms of the bulk gauge fields
under which they carry charge.  The various categories include the
O$p$-planes for $p=0,1,...8$ (which have the same supersymmetries and
the same kinds of charges as the D$p$-branes), the ONS5- and
OF1-planes which couple to the Neveu-Schwarz B-field, and the OP1
orientifold line which is not charged under a gauge field but does
carry momentum.

For the strongly coupled O6-plane, corrections from massive
Kaluza-Klein modes were important because the curvature radius $R$ of
the ten-dimensional metric (the `orientifold scale') was of the same
order as $R_{11}$ while other corrections (e.g., $\alpha'$
corrections) could be ignored (since e.g. $R \gg \ell_s$). To discover
which corrections may be important for the other O-planes, we should
similarly compare their geometric scales to the fundamental length
scales.

In the IIA cases, these fundamental 
scales are the string length $\ell_s$, the 10-d Planck
length $\ell_p^{(10)}$, and the scales of M theory: $R_{11}$ and the
11-d Planck length $\ell_p^{(11)}$.  While $R_{11}$ has no direct analogue in the IIB theory, distance
scales associated with the D-brane tensions lead to similar effects.
The fundamental scales are related through:
\begin{equation} \label{fund}
\ell_s = g^{-1/4} \ell_p^{(10)} = g^{-1/3} \ell_p \ \ \ \
\ell_p^{(10)} = (\ell_p^{(10)})^{9/8} R_{11}^{-1/8}, \ \ \ R_{11} = g \ell_s =
g^{2/3} \ell_p^{(11)}.
\end{equation}
For the (Ramond-Ramond) O$p$-planes, the scale is set by the D-brane
tension
\begin{equation} \label{tens}
T \sim \frac{m}{L^p} \sim \frac{1}{g \ell_s^{p+1}}.
\end{equation}
As a result, the radius of curvature $R$ of the 10-d metric for such
cases is
\begin{equation} \label{OR}
R^{7-p} \sim G_{10} T \sim g \ell_s^{7-p}.
\end{equation}

Consider first weak coupling. Then the string length is the largest
scale, and this is at least as big as the orientifold scale except for
$p=8$. Thus for $p \neq 8$, stringy corrections can invalidate the
pure supergravity description. Since these orientifolds preserve half
the supersymmetry, their interactions may be protected from
perturbative corrections.  However, we saw explicitly that
non-perturbative corrections arise in the O6 case.  Dualities then
imply that some corrections should be present for the other
O$p$-planes as well.  The O8-plane is an exception to this general
picture; it can have no significant corrections at weak coupling and
requires some new mechanism to protect the second law. The resolution
of this case is discussed in section \ref{o8} below.

At strong coupling, the situation is reversed and $R/\ell_s \sim
g^{1/(7-p)}$ is large for $p < 7$, so that $O(\alpha')$ corrections are
negligible for these cases. We now need to consider the other
fundamental scales, which are larger than the string scale at strong
coupling. We can rewrite (\ref{OR}) as 
\begin{equation}
R^{7-p} \sim g^{(p-3)/4} [\ell_p^{(10)}]^{7-p},
\end{equation}
so that $R/\ell_p^{(10)}$ is also large at strong coupling and the
associated corrections are small for $p > 3$. 
In terms of the 11-d Planck length
we have
\begin{equation}
R^{7-p} \sim g^{(p-4)/3} \ell_p^{7-p},
\end{equation}
so that 11-d quantum corrections are important for $p=4$. Thus, the
only cases where quantum corrections are not important are $p=6$,
considered in the previous section, and $p=5$. We will discuss $p=5$
in section \ref{o5} below, and argue that it receives 
non-perturbative corrections from D-string effects that are similar to those
described in section 2 for the orientifold 6-plane. 

There are also a few NS-type O-planes to consider. For the ONS5,
$G_{10} T \sim \ell_s^2$, and string corrections are always important.
For the OF1, we have $G_{10} T \sim g^2 \ell_s^6 \sim g^{1/2}
\ell_p^{(10)}{}^6 \sim \ell_{pl}^6$.  Thus, in the IIA theory, 11-d Planck corrections
are always important while in IIB string corrections are important
at weak coupling.   At strong coupling in the IIB theory, the D-string scale
$\ell_{D1} = g^{1/2} \ell_s$ (so that $T_{D1} \sim \ell_{D1}^{-2}$) becomes
large enough that we have
\begin{equation}
G_{10} T_{OF1} \sim g^2 \ell_s^6 = g^{-1} \ell_{D1}^6 \ll \ell_{D1}^6,
\end{equation}
so that D-string corrections will be important.  The T-dual OP1 has a
fixed line along a compact direction of radius $R_{compact}$, and a
tension $G_{10} T \sim g^2 \ell_s^8/R_{compact}^2$, which is smaller
than the OF1 tension for $R_{compact} > \ell_s.$ As a result, either
the same corrections are important as in the OF1 case above or
$R_{compact} \le \ell_s$ and string scale corrections must again be
considered.

Thus, a scaling analysis suggests that for any orientifold but the
O8-plane the spacetime description would receive corrections analogous
to the discussion in section \ref{o6} for the orientifold 6-plane.
For the O5-plane, the only relevant corrections should arise at the
scale of the D-string tension and we should be able to find them by
instanton methods in a probe brane calculation analogous to that of
section \ref{probe}.  We do so in section \ref{o5} below and then
proceed to analyze the case of the O8-plane in section \ref{o8}.

\subsection{Nonperturbative corrections and 
the O5-plane}
\label{o5}

Let us now turn to the strongly coupled O5-plane.  Our discussion
below should apply to any of the variants of the O5-plane, as
they differ only by discrete Neveu-Schwarz and Ramond-Ramond fluxes.
While string and (ten-dimensional) Planck scale corrections are small,
the effective description of strongly-coupled IIB is in fact the
S-dual weakly-coupled IIB.  Recall that there were no uncorrected IIB
orientifolds at weak coupling. In fact, the Ramond-Ramond O$p$-plane
is S-dual to the ONS5-plane, which had the potential for large stringy
corrections at weak coupling.  From the original strongly-coupled
point of view, these should arise somehow from the D-strings which are
becoming light.  This shows that such corrections are allowed by a
scaling analysis and our discussion of the second law predicts that
they will be present despite the supersymmetry of the situation.

How do we verify the existence of these corrections? We recall that the
non-perturbative corrections for the O6-plane can be seen in the
worldvolume field theory of probe branes.  A test D2-plane moving in
the O6 background is described by an $SU(2)$ gauge theory on the D2, and
the Atiyah-Hitchin metric is the associated moduli space metric.  The
massive Kaluza-Klein excitations in the Atiyah-Hitchin metric 
are related to instanton corrections to the corresponding moduli space.

We now attempt to repeat this analysis for the D1/O5-system.  The
relevant theory is now $SU(2)$ gauge theory in 2 Euclidean dimensions.
Since the topology of infinity is $S^1$, the space of Euclidean
solutions has instanton sectors associated with the winding of $U(1)$
subgroups around this $S^1$.  The lowest action configuration in this
sector should be a smooth Euclidean solution.  Indeed, a massive
version of the theory broken to a $U(1)$ subgroup is the abelian Higgs
model whose vortex instantons are well-known.

In two Euclidean dimensions, the action of such vortex instantons is in
general logarithmically divergent.  Because our fields are massless,
we expect this to be the case here.  Nonetheless, such
instantons can still contribute to the partition function.  The point
is that the action of an instanton/anti-instanton pair will be finite,
as the separation $s$ of the pair will cut off the logarithmic
divergence.  Such pairs therefore have an action of size $S \sim
\frac{1}{g} \ln s,$ where we have explicitly indicated the fact that
the action of any D-brane solution is proportional to $\frac{1}{g}$.
On the other hand, the volume of phase space available for a pair with
separation $s$ is proportional to the circumference of a circle with
radius $s$.  As a result, such instanton/anti-instanton pairs have an
entropy that diverges as $\ln s$, without an the accompanying factor of
$\frac{1}{g}$.  Note that at large $s$ the pair will provide an
approximate solution to the Euclidean equations of motion.  It is
therefore clear that, for large enough $g$, instanton/anti-instanton
pairs with large $s$ will provide important instanton contributions to
the partition function.  What we have just described is of course just
the usual Kosterlitz-Thouless phase transition \cite{KT} of two-dimensional
Euclidean systems.

This analysis verifies our prediction and shows that the moduli space for
D-strings moving in the O5-background receives corrections from instantons.  This
result and the analogy with the O6 system strongly suggests that we 
think of massive fields at scales set by the D1 tension as being
excited near an O5-plane.  Presumably, these fields carry negative
energy and affect the entropy of black holes in much the same way as in the O6 case
studied in section \ref{o6}.

\subsection{The O8-plane}
\label{o8}

Based on the O6-case, we generally expected non-trivial corrections to
the (10d) supergravity description of the orientifolds. This is borne
out, with the sole exception of the O8-plane at weak coupling.  For this
case, all of the corrections considered above are
negligible. Furthermore, as it has co-dimension one, the O8-plane will
attract a positive-mass black hole, so this case is also an exception to
the argument in the introduction about kinetic energy being required to impel
a black hole toward a negative tension brane. Since all the
corrections are suppressed, the resolution of these puzzles must lie
within classical (massive) type IIA supergravity.

Let us consider first the question of what happens when an event horizon encounters an O8-brane.
We would like to suggest that the resolution of our thermodynamic concerns is related
to a well-known difficulty in dealing with O8-planes and the related
D8-branes.  The solution corresponding to any of these objects in
isolation has a singularity at a finite distance of order
$\frac{1}{g}\ell_s$ from the brane.  The usual interpretation of this
singularity is that it is impossible to separate an O8- or D8-brane
from an oppositely charged such brane by more than a distance of order
$\frac{1}{g}\ell_s$.  Thus, when we consider an O8-plane, we should
remember that there must necessarily be nearby either an anti O8-plane (in which
case the entire universe must be small as these are world-ending
branes) or a D8-brane (which has positive tension).  As a result,
there is simply no setting where one may ask about a black hole (or in fact a black $p$-brane) which
begins farther away from the O8-plane than the O8 curvature radius $R \sim \frac{1}{g}\ell_s$
and collides with the negative
tension orientifold without at the same time encountering a similar
amount of positive energy material.

It is instructive to briefly consider black $p$-branes that begin close to
the orientifold.  Such a black hole will act much like its cousin in empty
space if the size $R_{bb}$ of the black $p$-brane is much smaller than the
O8-plane length scale $R_{O8} = \frac{1}{g}\ell_s$.  If
$R_{bb}/R_{O8}$ is non-negligible, we expect that distortions of this
order will be present in the black brane horizon and consequently that
the area $A$ of such a black brane will differ from the area $A_0$ of its flat space
cousin by an amount of order $A_0 \frac{R_{bb}}{R_{O8}}$.  Furthermore, the exact
distortion and the corresponding effect on the horizon area will
depend on the precise location of the black brane, so that the horizon
area of a black brane colliding with the O8-plane will change during the collision by an
amount of this order in a manner that is beyond the scope of this work to predict.

Let us also estimate the magnitude of the supposed entropy reduction
due to the collision of a black $p$-brane with the O8-brane. The
amount of negative tension absorbed by the brane is roughly
proportional to the tension $T_{O8}$ of the orientifold times the
cross sectional area $R_{bb}^{8-p}$ of the black brane, whereas the
original tension of the black brane is of size
$\frac{R_{bb}^{7-p}}{G_{10}}$, where $G_{10}$ is the ten-dimensional
Newton's constant.  Since $G_{10}T_{O8} \sim 1/R_{O8}$, we have a
fractional change in tension of order $\Delta T/T \sim R_{bb}/R_{O8}$
and we would expect a fractional decrease in its area of the same
order. But this is of the same size as the dynamical distortion term
caused by the proximity of the O8-plane! These finite-distance
corrections can therefore resolve the potential decrease in area.  We
suggest that this is exactly what happens, though it is difficult to
study the effect in detail.

In the introduction, we argued that the total mass of the black hole
typically could not be reduced by collision with a negative-energy
source due to the repulsive gravitational potential. In the present
case, this argument breaks down, because the O8-brane is of
co-dimension one. However, this problem contains its own solution:
since the O8-brane is co-dimension one while black branes always have
co-dimension greater than one, a black $p$-brane cannot engulf the
entire O8-brane.  Instead, part of the O8 necessarily remains in the
spacetime to source significant distortions of the black $p$-brane.
Hence, although the mass of the black hole may decrease, one cannot
conclude that the black hole's area decreases when it collides with
the O8-brane.  Note also that the above scaling analysis suggests that
a black hole cannot engulf enough O8 negative tension to bring its
total mass to zero or below.

\section{AdS Soliton}
\label{soliton}

A final 
negative-energy object in string theory is the AdS
soliton~\cite{hor-myers}. 
Like the Atiyah-Hitchin metric, the AdS soliton is a pure gravitational solution and is
not particularly stringy in and of itself.  However, it is present in
string theory and its construction was motivated by stringy considerations.  
As a result, this object is as connected with the consistency of string theory and the second law
as the objects that we have previously discussed.

The AdS soliton is an asymptotically AdS solution of the vacuum
Einstein equations with a cosmological constant and arises in
string theory in the context of sphere compactifications. The metric
for the AdS soliton in $p+2$ dimensions is~\cite{hor-myers}
\begin{equation} \label{solm}
ds^2 = {r^2 \over l^2} \left[ \left( 1 - {r_0^{p+1} \over r^{p+1}}
\right) d\tau^2 + (dx^i)^2 - dt^2 \right] + \left( 1 - {r_0^{p+1}
\over r^{p+1}} \right)^{-1} {l^2 \over r^2} dr^2,
\end{equation}
where $i=1\ldots p-1$, $l$ is the cosmological length scale, and the
coordinate $\tau$ is compactified with period
\begin{equation}
\beta = {4\pi l^2 \over (p+1) r_0} .
\end{equation}
We also take the $x^i$  to be compactified on a $p-1$ dimensional torus of volume $V_{p-1}$.
This is a smooth metric, but it has negative energy, 
\begin{equation}
E = -{r_0^{p+1} \beta V_{p-1} \over 16 \pi G_{p+2} l^{p+2}} ,
\end{equation}
where a definition of $E$ has been used in which the energy of AdS space vanishes.

If one dimensionally reduces along $\tau$, this
looks like a $p+1$ dimensional spacetime with a negative-energy
`brane' at $r=r_0$.  Now, for the metric (\ref{solm}) the proper 
length of the $\tau$ circle diverges
at large $r$ so that $p+2$ dimensional physics is relevant at any value of $\beta$.
On the other hand, one can make the physics truly $p+1$ dimensional by cutting of the spacetime
at large $r$ as in \cite{LMW}.  A collision with a black hole in the $p+1$
dimensional spacetime is then described by considering a black string
wrapping the $\tau$ direction.

This example differs from the previous cases in that
the amount of negative energy is a free parameter (controlled by $\beta$), 
so that it would be difficult to
resolve our concerns using fundamental corrections to the
supergravity description. In addition, like the O8-brane, the AdS
soliton is co-dimension one when considered as a brane and one may check directly
that it is gravitationally attractive.

Remarkably, the brane at $r=r_0$ {\it cannot} cross the black hole
horizon in the usual sense. Since $r=r_0$ is the origin of the
$r,\tau$ plane in the $p+2$ dimensional geometry, an $S^1$ of horizon
generators would go to zero size if the black string horizon crossed
$r=r_0$.  Since this geometry is smooth, such a caustic cannot occur
on an event horizon \cite{WaldBook,HE}. However, the $p+1$ dimensional
black hole is assuredly attracted to the negative-energy `brane'. So
what happens? The only possibility is that the brane enters the
horizon at a past endpoint, where new generators are entering the
horizon and the black hole is still forming.  One can see that this is
so by again using the rotational symmetry.  The point $r=r_0$ cannot
cross the event horizon along a null generator, as there is no
distinguished direction for the corresponding null ray to be
travelling.  Since the negative-tension brane does not intersect a
pre-existing family of horizon generators, the argument in the
introduction is simply inapplicable. Once again, in the
$p+2$-dimensional description, the geometry is everywhere smooth so
that the horizon area will not decrease. In particular, the final
horizon area must be greater than any reasonable definition of an
`initial area.'

Note that in contrast to recent work on higher-dimensional
solutions~\cite{torus}, we don't expect the horizon of our $p+2$
``black string'' to be toroidal in any invariant sense; the picture
here is more like~\cite{shapiro}, where one forms a horizon with
initially toroidal spacelike slices, but the hole closes up faster
than the speed of light.  In the 3+1 analogue \cite{shapiro}, there
are no causal curves linking the torus\footnote{In our higher
dimensional setting this question is moot as tori of our dimension
link with surfaces and not curves.}  which extend to infinity (in
agreement with topological censorship \cite{TopCen}). The cosmological
constant (i.e., the $r^2/l^2$ factor in $g_{\tau \tau}$) will
accelerate the wrapped black brane toward $r=r_0$ even if we place the
string far away initially, so the `donut hole' in the middle can close
up before observers can pass through it.

Finally, we note an interesting property of this ``brane'' that is
intimately associated with the resolution above.  Because in $p+2$
dimensions the $\tau$ circle shrinks to zero size at $r=r_0$, the
$p+1$ dimensional Einstein frame metric necessarily degenerates at the
location of the brane.  Such a singularity of the $p+1$ metric then
forces one to embed the description of this brane in some theory that
is more complete than just $p+1$ dimensional Einstein-Hilbert gravity
in order to describe the brane.  Thus we have once again failed to
find a stringy negative tension brane which couples only to
Einstein-Hilbert gravity.

\section{Discussion}

We have addressed a number of stringy negative tension branes and, in
each case, we have identified a plausible mechanism that could make these
objects compatible with the generalized second law.  We have provided
various pieces of evidence, including scaling analyses, perturbative
treatments, finding non-perturbative corrections to moduli spaces, 
and analyses of black hole event horizons.  

The two cases of the O8-plane and the AdS soliton were exceptional.
For the O8-plane we argued that it is only the required proximity to
positive tension D8-branes and the distortions these objects
induce in nearby black holes that allows the second law to hold.
It would clearly be of interest to compute such distortions, perhaps
in the perturbative regime of a small black hole, in order to verify
our conjecture that the distortions increase the surface area.
On the other hand, we argued for the AdS soliton that it was simply 
impossible for this brane to meet a ten-dimensional event horizon
except at the initial events where the horizon first forms and null
generators are added.

In most of the orientifold cases, it was possible to identify a set of
corrections that should significantly modify the na\"\i ve picture of
an infinitely thin negative energy brane interacting with simple
(Einstein-Hilbert like) gravitational degrees of freedom.  In the O6
case we have seen that these corrections have the interpretation of
resulting from massive fields for which the orientifold acts as a
source, such that the massive fields have order-one excitations near
the orientifold.  We believe that this picture holds in general, and
the instanton corrections identified for the O5-plane are consistent
with this conjecture.  These massive fields should have curvature
couplings to the na\"\i ve Einstein metric, so that in terms of this
metric the entropy of a black hole will not be given exactly by
$\frac{1}{4}$ of the area in Planck units.  Such corrections to the
entropy should be relevant when an orientifold encounters the horizon
of the black hole.

If the
black hole completely engulfs the orientifold then these corrections should
die out as the system equilibrates.  However, for such cases
the orientifold has co-dimension greater than 1 and, 
as described in the introduction, energy considerations suggest that
second law violations are of concern only during the collision and not 
in the final equilibrium state.  It is therefore fitting that
our resolution should involve corrections of a similar transitory nature.

When we spoke of corrections above, we generally mean ways in which a
spacetime description of an orientifold would differ from that
obtained by analytic continuation from the positive tension D-branes
to appropriate negative tensions.  Although we argued that such
corrections should exist, we did not map out these corrections in
detail for any case other than the O6-plane.  Some progress was made
for the O5-plane, where we saw that the corrections were related to
the Kosterlitz-Thouless unbinding of instanton/anti-instanton pairs.

It would be interesting to find the explicit form of these corrections
in other cases as well.  Recall, however, that our interest is
generally in the region close to the orientifold where we expect the
corrections to dominate over the negative tension D-brane geometry.
In contrast, natural methods to compute these effects do so in the
regime where the corrections are small.  While such calculations are
therefore unlikely to produce fully satisfactory results, it may be
that they reveal interesting properties as we saw in the perturbative
calculations of sections \ref{10d} and \ref{mass-stress}.

The discussion of energy and entropy above was carried out using the
na\"\i ve Einstein frame.  The reader may wonder if one can do better by
correcting the metric to obtain a `true Einstein frame.'  For
definiteness, let us return to the case of the O6-plane and the
Atiyah-Hitchin metric in 11-dimensions.  In ten dimensions, one may
try to find a true Einstein frame by correcting the definition
(\ref{kkred}) of the ten-dimensional Einstein metric at each order in
$h_{ab}$.  That this is in fact possible may be seen from the fact
that the curvature couplings result from the expansion of the
11-dimensional Ricci tensor and are thus linear the ten-dimensional
Ricci tensor -- the terms do not involve higher powers of the Ricci
tensor or other components of the ten-dimensional Riemann tensor.
Such terms can always be cancelled by the change in the 10-dimensional
Einstein term (essentially the Einstein tensor contracted with the
modification of $g^{E,10}_{ab}$) induced by an appropriate
modification of the metric.

As a result, a true Einstein frame will exist at each perturbative
order.  Since black hole entropy will be given by the area in this
true Einstein frame, this area must match the area as measured by the
eleven-dimensional metric.  Thus, to the extent that one works in such
a true Einstein frame, one expects the description of the physics to
match that given in eleven dimensions.  One should therefore not find
any localized negative energy at all and, to the extent that such a
frame makes sense non-perturbatively, the geometry should close off
smoothly at the origin with no singular brane to mark any particular
points as special.  One expects that from this perspective the
negative energy is again purely a consequence of the boundary
conditions.  However, in the weak coupling ($g \rightarrow 0$) limit
all of this structure shrinks to zero size and one is left with an
object effectively described as a negative tension brane with orbifold
boundary conditions so long as one does not probe this brane too
closely.

Note that the `true Einstein metric' described above involves much
more than just a conformal rescaling of the metric.  In contrast to
the more familiar relation between the string and Einstein frames in
massless type II supergravity, the present change to the true Einstein
frame field will therefore introduce a new notion of null rays, causal
structure, and black hole horizons.

Interestingly, in no case did we find a resolution in a regime where
the negative tension brane could be described in terms of couplings
to pure Einstein-Hilbert gravity.  For most of the orientifolds, extra
massive fields with complicated curvature couplings seemed to be required
for thermodynamic stability.  While the O8-brane did not require
massive fields in quite the same sense, its resolution resulted
from the couplings of massive type IIA supergravity which force
the solution to become singular at a finite distance from an isolated brane.
When the AdS soliton was viewed as a 
negative tension brane, the metric necessarily degenerated at the brane
and thus passed out of the regime of pure gravity.   These
are in sharp contrast to the rather benign looking phenomenological
negative tension branes used in e.g. \cite{RS}, for which the interaction
of the metric and brane can be described in terms of the Israel
junction conditions.  It appears that string theory gives no
motivation for stability of simple phenomenological negative tension
branes coupled to Einstein-Hilbert like gravity. 

While we have by no means ruled out further mechanisms to enforce
the second law, it is far from clear what these might be.
As a result, one suspects that generic phenomenological negative
tension branes face difficulties with the second law and, even when
placed at an orbifold, are likely
to experience instabilities involving black holes of the sort discussed
in \cite{MT,MRST}.  While some more complete treatment will
therefore eventually be required, the simplicity of 
such phenomenological negative tension branes will no doubt maintain
their important role in exploring novel and interesting new applications
of negative energy objects.  

\acknowledgments

We would like to thank Mark Bowick, Clifford Johnson, Dan Kabat, Mark
Trodden, and Arkady Tseytlin for useful discussions and to acknowledge
the support of the Aspen Center for Physics, where the initial stages
of this work were completed.  D.M. was supported in part by NSF grant
PHY00-98747 and by funds from Syracuse University.  S.F.R. is
supported by an Advanced Fellowship from EPSRC. 
S.F.R. also acknowledges the support of the Isaac Newton Institute for
Mathematical Sciences during the completion of this work.


\begin{thebibliography}{99}

\bibitem{Sag} A. Sagnotti, ``Open Strings and their Symmetry Groups,''
ROM2F-87/25, {\it Talk presented at the Cargese Summer Institute on 
Non-Perturbative Methods in Field Theory, Cargese, Italy, Jul 16-30, 
1987}.


\bibitem{H} P. Ho\v{r}ava, ``Strings on World Sheet Orbifolds,''
Nucl. Phys. {\bf B327} 461 (1989).
%%CITATION = NUPHA,B327,461;%%

\bibitem{Joe}  J. Polchinski, {\it String Theory,} (Cambridge U. Press 1998).

\bibitem{GP} E. G. Gimon and J. Polchinski, ``Consistency Conditions
for Orientifolds and D-Manifolds,'' Phys. Rev. D {\bf 54}  1667 (1996),
hep-th/9601038.
%%CITATION = HEP-TH 9601038;%%

\bibitem{RS}
L.~Randall and R.~Sundrum,
``A large mass hierarchy from a small extra dimension,''
Phys.\ Rev.\ Lett.\ {\bf 83}, 3370 (1999),
hep-ph/9905221.
%%CITATION = HEP-PH 9905221;%%

\bibitem{WaldBook} R. M. Wald, {\it General Relativity} (Chicago, 1984).

\bibitem{AGM}
P.~S.~Aspinwall, B.~R.~Greene and D.~R.~Morrison,
``Space-time topology change: The Physics of Calabi-Yau moduli
space,'' hep-th/9311186.
%%CITATION = HEP-TH 9311186;%%

\bibitem{Juan}
O.~Aharony, S.~S.~Gubser, J.~Maldacena, H.~Ooguri and Y.~Oz,
``Large N field theories, string theory and gravity,''
Phys.\ Rept.\  {\bf 323}, 183 (2000),
hep-th/9905111.
%%CITATION = HEP-TH 9905111;%%

\bibitem{ISMY}
N.~Itzhaki, J.~Maldacena, J.~Sonnenschein and S.~Yankielowicz,
``Supergravity and the large N limit of theories with sixteen
supercharges,'' 
Phys.\ Rev.\ D {\bf 58}, 046004 (1998),
hep-th/9802042.
%%CITATION = HEP-TH 9802042;%%

\bibitem{EGJ} H. Epstein, V. Glaser, and A. Jaffe, ``Nonpositivity of
  the energy density in quantized field theories,'' Nuovo Cim. {\bf 36}
(1965) 1016.

\bibitem{Kuo} C.~I.~Kuo,
``A revised proof of the existence of negative energy density in
quantum  field theory,'' 
Nuovo Cim.\ B {\bf 112}, 629 (1997),
gr-qc/9611064.
%%CITATION = GR-QC 9611064;%%

\bibitem{Tipler} F. J. Tipler, ``Energy conditions and spacetime
  singularities,'' Phys. Rev. D {\bf 17},  2521 (1978).
%%CITATION = PHRVA,D17,2521;%%

\bibitem{Romans1} T.~A.~Roman,
``Quantum Stress Energy Tensors And The Weak Energy Condition,''
Phys.\ Rev.\ D {\bf 33}, 3526 (1986).
%%CITATION = PHRVA,D33,3526;%%

\bibitem{FR1} L.~H.~Ford and T.~A.~Roman,
``The quantum interest conjecture,''
Phys.\ Rev.\ D {\bf 60}, 104018 (1999),
gr-qc/9901074.
%%CITATION = GR-QC 9901074;%%

\bibitem{Borde} A. Borde, ``Geodesic focusing, energy conditions and
  singularities,'' Class. Quant. Grav. {\bf 4}, 343 (1987).

\bibitem{MorTh} M.~S.~Morris and K.~S.~Thorne,
``Wormholes In Space-Time And Their Use For Interstellar Travel: A
  Tool For Teaching General Relativity,'' 
Am.\ J.\ Phys.\  {\bf 56}, 395 (1988).
%%CITATION = AJPIA,56,395;%%

\bibitem{MTY} M.~S.~Morris, K.~S.~Thorne and U.~Yurtsever,
``Wormholes, Time Machines, And The Weak Energy Condition,''
Phys.\ Rev.\ Lett.\  {\bf 61}, 1446 (1988).
%%CITATION = PRLTA,61,1446;%%


\bibitem{Alc} M.~Alcubierre,
``The warp drive: hyper-fast travel within general relativity,''
Class.\ Quant.\ Grav.\  {\bf 11}, L73 (1994),
gr-qc/0009013.
%%CITATION = GR-QC 0009013;%%

\bibitem{Olum} K.~D.~Olum,
``Superluminal travel requires negative energies,''
Phys.\ Rev.\ Lett.\  {\bf 81}, 3567 (1998),
gr-qc/9805003.
%%CITATION = GR-QC 9805003;%%

\bibitem{GW} S.~Gao and R.~M.~Wald,
``Theorems on gravitational time delay and related issues,''
Class.\ Quant.\ Grav.\  {\bf 17}, 4999 (2000),
gr-qc/0007021.
%%CITATION = GR-QC 0007021;%%

\bibitem{FR2} L.~H.~Ford and T.~A.~Roman,
``Moving Mirrors, Black Holes And Cosmic Censorship,''
Phys.\ Rev.\ D {\bf 41}, 3662 (1990).
%%CITATION = PHRVA,D41,3662;%%

\bibitem{FR3} L.~H.~Ford and T.~A.~Roman,
``'Cosmic flashing' in four-dimensions,''
Phys.\ Rev.\ D {\bf 46}, 1328 (1992).
%%CITATION = PHRVA,D46,1328;%%

\bibitem{Ev} A. Everett, ``Warp drive and causality,'' Phys. Rev. D
  {\bf 53},  7365 (1996). 
%%CITATION = PHRVA,D53,7365;%%

\bibitem{HE} S. W. Hawking and G. F. R. Ellis, {\it The large scale
structure of space-time} (Cambridge U. Press, 1973).

\bibitem{AE} A. Einstein, as quoted in 
M.J. Klein, ``Thermodynamics in Einstein's Universe", Science, {\bf 157}
(1967) 509.

\bibitem{MT}
D.~Marolf and M.~Trodden,
``Black holes and instabilities of negative tension branes,''
Phys.\ Rev.\ D {\bf 64}, 065019 (2001),
hep-th/0102135.
%%CITATION = HEP-TH 0102135;%%


\bibitem{SUSY}
R.~Altendorfer, J.~Bagger and D.~Nemeschansky,
``Supersymmetric Randall-Sundrum scenario,''
Phys.\ Rev.\ D {\bf 63}, 125025 (2001),
hep-th/0003117.
%%CITATION = HEP-TH 0003117;%%

\bibitem{MRST} D. Marolf, J. Rozowsky, P. Silva, and M. Trodden, 
in preparation.

\bibitem{Pri} Private communication with the authors of \cite{MRST}.

\bibitem{wald}
V.~Iyer and R.~M.~Wald,
``A Comparison of Noether charge and Euclidean methods for computing
the entropy of stationary black holes,'' 
Phys.\ Rev.\ D {\bf 52}, 4430 (1995),
gr-qc/9503052;
%%CITATION = GR-QC 9503052;%%
V.~Iyer and R.~M.~Wald,
``Some properties of Noether charge and a proposal for dynamical black
hole entropy,'' 
Phys.\ Rev.\ D {\bf 50}, 846 (1994),
gr-qc/9403028.
%%CITATION = GR-QC 9403028;%%

\bibitem{hor-myers}
G.~T.~Horowitz and R.~C.~Myers,
``The AdS/CFT correspondence and a new positive energy conjecture for
general relativity,'' 
Phys.\ Rev.\ D {\bf 59}, 026005 (1999),
hep-th/9808079.
%%CITATION = HEP-TH 9808079;%%

\bibitem{LMW} 
F.~Leblond, R.~C.~Myers and D.~J.~Winters,
``Consistency conditions for brane worlds in arbitrary dimensions,''
JHEP {\bf 0107}, 031 (2001)
hep-th/0106140.
%%CITATION = HEP-TH 0106140;%%


\bibitem{Sen}
A.~Sen,
``A note on enhanced gauge symmetries in M- and string theory,''
JHEP {\bf 9709}, 001 (1997),
hep-th/9707123.
%%CITATION = HEP-TH 9707123;%%

\bibitem{Seiberg}
N.~Seiberg,
``IR dynamics on branes and space-time geometry,''
Phys.\ Lett.\ B {\bf 384}, 81 (1996),
hep-th/9606017.
%%CITATION = HEP-TH 9606017;%%

\bibitem{SW}
N.~Seiberg and E.~Witten,
``Gauge dynamics and compactification to three dimensions,''
hep-th/9607163.
%%CITATION = HEP-TH 9607163;%%

\bibitem{AHbook} M. F. Atiyah and N. Hitchin, {\it The geometry and
dynamics of magnetic monopoles} (Princeton U. Press, 1988).

\bibitem{HP} A.~Hanany and B.~Pioline,
``(Anti-)instantons and the Atiyah-Hitchin manifold,''
JHEP {\bf 0007}, 001 (2000),
hep-th/0005160.
%%CITATION = HEP-TH 0005160;%%

\bibitem{KT} J.~M.~Kosterlitz and D.~J.~Thouless,
``Ordering, Metastability And Phase Transitions In Two-Dimensional
Systems,'' 
J.\ Phys.\ C {\bf 6}, 1181 (1973).
%%CITATION = JPCBA,C6,1181;%%

\bibitem{BGS} O.~Bergman, E.~G.~Gimon and S.~Sugimoto,
``Orientifolds, RR torsion, and K-theory,''
JHEP {\bf 0105}, 047 (2001),
hep-th/0103183.
%%CITATION = HEP-TH 0103183;%%

\bibitem{Dorey}
N.~Dorey, V.~V.~Khoze, M.~P.~Mattis, D.~Tong and S.~Vandoren,
``Instantons, three-dimensional gauge theory, and the Atiyah-Hitchin
manifold,'' 
Nucl.\ Phys.\ B {\bf 502}, 59 (1997),
hep-th/9703228.
%%CITATION = HEP-TH 9703228;%%

\bibitem{monopole} G.~W.~Gibbons and N.~S.~Manton, ``Classical and
quantum dynamics of BPS monopoles,'' Nucl.\ Phys.\ B {\bf 274} (1986)
183.
%%CITATION = NUPHA,B274,183;%%

\bibitem{SY} 
R.~Schoen and S.~T.~Yau,
``On The Proof Of The Positive Mass Conjecture In General Relativity,''
Commun.\ Math.\ Phys.\  {\bf 65}, 45 (1979);
%%CITATION = CMPHA,65,45;%%
R.~Schoen and S.~T.~Yau,
``Proof Of The Positive Mass Theorem. 2,''
Commun.\ Math.\ Phys.\  {\bf 79}, 231 (1981).
%%CITATION = CMPHA,79,231;%%

\bibitem{WPM} E. Witten, 
E.~Witten,
``A Simple Proof Of The Positive Energy Theorem,''
Commun.\ Math.\ Phys.\  {\bf 80}, 381 (1981).
%%CITATION = CMPHA,80,381;%%


\bibitem{action}
G.~W.~Gibbons, S.~W.~Hawking and M.~J.~Perry,
``Path integrals and the indefiniteness of the gravitational action,''
Nucl.\ Phys.\ B {\bf 138}, 141 (1978).
%%CITATION = NUPHA,B138,141;%%

\bibitem{GL}
R.~Gregory and R.~Laflamme,
``Black strings and p-branes are unstable,''
Phys.\ Rev.\ Lett.\  {\bf 70}, 2837 (1993),
hep-th/9301052.
%%CITATION = HEP-TH 9301052;%%


\bibitem{torus}
R.~Emparan and H.~S.~Reall,
``A rotating black ring in five dimensions,''
hep-th/0110260.
%%CITATION = HEP-TH 0110260;%%

\bibitem{shapiro}
S.~L.~Shapiro, S.~A.~Teukolsky and J.~Winicour,
``Toroidal black holes and topological censorship,''
Phys.\ Rev.\ D {\bf 52}, 6982 (1995).
%%CITATION = PHRVA,D52,6982;%%


\bibitem{TopCen}
J.~L.~Friedman, K.~Schleich and D.~M.~Witt,
``Topological Censorship,''
Phys.\ Rev.\ Lett.\  {\bf 71}, 1486 (1993)
[Erratum-ibid.\  {\bf 75}, 1872 (1993)],
gr-qc/9305017.
%%CITATION = GR-QC 9305017;%%

    
\end{thebibliography}
\end{document}